\documentclass[aps,prl,twocolumn,preprintnumbers,superscriptaddress]{revtex4-1}

\usepackage[usenames,dvipsnames,table]{xcolor}
\usepackage{graphicx,amsmath,amssymb,amsthm,multirow,array,bm,bbm,esint}
\usepackage[mathscr]{eucal}
\usepackage[bbgreekl]{mathbbol}
\usepackage{epsf,amsfonts}
\usepackage{hyperref}

\newcommand{\beq}{\begin{equation}}
\newcommand{\eeq}{\end{equation}}
\newcommand\cnot[1]{%
  \mathrel{\ooalign{\hfil$#1$\hfil\cr\hfil$/$\hfil\cr}}}

\begin{document}

\title{Chaos in AdS$_2$ holography}

\author{Kristan Jensen}
\email{kristanj@sfsu.edu}
\affiliation{Department of Physics and Astronomy, San Francisco State University, San Francisco, CA 94132}

\date{\today}

\begin{abstract}

We revisit AdS$_2$ holography with the Sachdev-Ye-Kitaev models in mind. Our main result is to rewrite a generic theory of gravity near an AdS$_2$ throat as a novel hydrodynamics coupled to the correlation functions of a conformal quantum mechanics. This gives a prescription for the computation of $n$-point functions in the dual quantum mechanics. We thereby find that the dual is maximally chaotic.

\end{abstract}


\maketitle

\textit{Introduction.}~The Sachdev-Ye-Kitaev (SYK) models~\cite{Sachdev:1992fk,kitaev} are quantum mechanical systems with random all-to-all interactions. They have been recently argued to have a gravity dual in two dimensions.

The basic SYK model is a theory of $2N$ Majorana fermions $\psi^a$ ($a=1,..,2N$) perturbed by quenched disorder. The Hamiltonian is
\beq
\label{E:SYKh}
H =\sum_{a,b,c,d} \frac{J_{abcd}}{4!}\psi^a \psi^b\psi^c\psi^d\,,
\eeq
where $\overline{J_{abcd}}=0$ and $\overline{J_{abcd}J^{abcd}}= 3!J^2/(2N)^3$. At a temperature $T$, there is a single dimensionless coupling $J/T$. The high-temperature theory has $2N$ weakly interacting fermions, while the low-temperature theory is strongly correlated. Crucially, the theory is soluble at large $N$ (see e.g.~\cite{Polchinski:2016xgd,Maldacena:2016hyu}).

There are two main pieces of evidence that the SYK models have a gravity dual. The first is an emergent conformal symmetry at low energies, together with a large $N$ extremal entropy~\footnote{Sachdev~\cite{Sachdev:2015efa} has also noted a relation between the ``spectral asymmetry'' and the extremal thermodynamics in the Sachdev-Ye models. That relation also holds in AdS$_2$ holography. However this relation is a consequence of the emergent conformal symmetry~\cite{HJ,Davison:2016ngz}.}. The second is much more non-trivial. The SYK models saturate the ``chaos bound'' of~\cite{Maldacena:2015waa} on the Lyapunov exponent, which characterizes the rate of growth of certain out-of-time-ordered four-point functions~\cite{1969JETP...28.1200L,Shenker:2013pqa}. This bound, which exists in any quantum system, is $2\pi T$. Conformal field theories with an Einstein gravity dual also saturate the chaos bound~\cite{Shenker:2013pqa}, which led Kitaev~\cite{kitaev} to conjecture that the SYK model gives a toy model for quantum gravity in two dimensions (see also~\cite{PhysRevLett.105.151602,Sachdev:2015efa}). 

This prospect brings us back to the AdS$_2$/CFT$_1$ correspondence, along with all of its baggage. The AdS$_2$/CFT$_1$ correspondence has never been satisfactorily developed, largely due to problems on both sides of a putative duality. In one dimension, field theories are ordinary quantum mechanics, and so we hereafter refer to a CFT$_1$ as a conformal quantum mechanics (CQM)~\footnote{This is a rather different sense of conformal invariance than that in the quantum mechanics of a particle in an inverse-square potential (see e.g.~\cite{deAlfaro:1976vlx,Chamon:2011xk}), whose Hilbert space decomposes into irreducible representations of the global conformal group $SL(2;\mathbb{R})$.}.

On the CQM side, one runs into a paradox due to Polchinski~\cite{Jensen:2011su}. Let $\rho(E)$ be the density of states. Scale invariance implies
\beq
\rho(E) = e^{S_0}\delta(E) + \frac{e^{S_1}}{E}\,.
\eeq
If the second term is nonzero, then there must be an infrared cutoff $\Lambda_{\rm IR}$, but if it vanishes then a CQM is a topological theory with no dynamics.

This CQM paradox is dual to the fact that AdS$_2$ spacetimes cannot support finite-energy excitations. Injecting a lump of energy into an AdS$_2$ throat leads to strong backreaction, which cannot be consistently analyzed within the throat.

These two paradoxes are dual to each other in that they reflect modest UV/IR mixing. On the CQM side, ``irrelevant deformations'' to the density of states allow for consistent time evolution and non-topological correlators, while on the gravity side AdS$_2$ throats do not admit a decoupling limit. A consistent study of scattering requires the flow to the throat.

There is a connection to large $N$ limits here, in that these paradoxes arise at finite $N$. In the strict $N\to\infty$ limit, there is nothing wrong with a generalized free CQM~\cite{ElShowk:2011ag}, while backreaction disappears on the gravitational side. However, at finite $N$, there is no such thing as an interacting CQM or AdS$_2$ holography. A large $N$ theory may be only approximately conformally invariant, with conformal invariance broken at $\mathcal{O}(1/N)$, as advocated in~\cite{Almheiri:2014cka}.

The point of this Letter is two-fold. First, to assess the viability of an SYK/AdS$_2$ correspondence. Second, to revisit AdS$_2$ holography. For theories dual to dilaton gravity with an AdS$_2$ near-horizon, we derive an effective hydrodynamic action for the near-AdS$_2$ physics from which we see that they saturate the chaos bound.

\emph{Note}: While this Letter was nearing completion, Maldacena and Stanford posted a very interesting paper~\cite{Maldacena:2016hyu} which has some overlap with this work.

\textit{The SYK models.}~The theory of $2N$ Majorana fermions $\psi^a$ ($a=1,..,2N$)~\footnote{The theory of an odd number of Majorana fermions has a global gravitational anomaly~\cite{AlvarezGaume:1983ig}: its partition function does not have definite sign, due to an ambiguity in the square root of the determinant of the Dirac operator. Relatedly, a naive computation of the thermal entropy of a single Majorana fermion gives $S = \ln \sqrt{2}$, rather than the logarithm of an integer.} is an exact CQM:
\beq
S_{\psi} = \sum_a\int dt \, \psi_a \partial_t \psi^a\,.
\eeq
It is merely a system of $2^N$ zero-energy states, and so has a large extremal entropy $S = N \ln 2$. The two-point function of $\psi$ is topological,
\beq
\langle \psi^a(t) \psi^b(0) \rangle = \frac{\delta^{ab}}{2}\text{sgn}(t)\,.
\eeq

This theory admits $N $ relevant deformations built from fermion monomials. The SYK model is the theory of $2N$ fermions perturbed by quenched disorder for the quartic monomial,
\beq
S_{SYK} = \int dt \big(  \sum_a\psi_a\partial_t \psi^a - \sum_{a,b,c,d}\frac{J_{abcd} }{4!}\psi^{a} \psi^b\psi^c \psi^{d} \big) \,.
\eeq
Note that if the source $J_{abcd}$ was not disordered, then the four-Fermi interaction would break the global symmetry. However, the quenched disorder preserves the full $SO(2N)$ flavor symmetry of the free-field fixed point.

The SYK model realizes an emergent conformal symmetry at low energies and large $N$. At $T=0$, the solution to the leading large $N$ Schwinger-Dyson equation for the two-point function of $\psi^a$ is
\beq
\label{E:SYK2pt}
\langle  \psi^a(t) \psi^b(0) \rangle  =\left(  \frac{1}{4\pi J^2}\right)^{1/4}\frac{\text{sgn}(t)\delta^{ab}}{|t|^{1/2}}\,, \quad t \gg 1/J\,,
\eeq
so that $\psi^a$ has dimension $1/4$ in the infrared.

There is a generalization of the SYK model characterized by two integers, the number of fermions $2N$ and the degree $q$ of the disordered interaction:
\beq
S_q = S_{\psi} -  \int dt \sum_{a_1,..,a_q}\frac{J_{a_1..a_q}}{q!}\psi^{a_1}..\psi^{a_q}\,.
\eeq
This theory also hosts an emergent conformal symmetry at low energies and large $N$ (with $q\ll N$) where $\psi^a$ behaves like a dimension-$1/q$ operator in the IR. 

The SYK model exhibits another hallmark of emergent conformal symmetry in one dimension: it has a large $N$ extremal entropy. Standard large $N$ power counting shows that the leading contribution to the low-temperature, large $N$ thermal partition function is the one-loop determinant of the inverse, resummed fermion propagator~\eqref{E:SYK2pt}. Conformally mapping to the thermal circle, the thermal Euclidean two-point function of $\psi$ is
\beq
G(\omega_n) \propto \frac{\Gamma \left(  \Delta -n+\frac{1}{2} \right)}{\Gamma\left( 1 - \Delta - n + \frac{1}{2}\right)}\,,\qquad \Delta = \frac{1}{q}\,,
\eeq
where $\omega_n = 2\pi (n - 1/2)T$ is the $n^{th}$ Matsubara frequency. The extremal entropy is given by
\beq
\frac{S}{N} =  \sum_n \ln |G^{-1}(\omega_n)| + \mathcal{O}(N^{-1})\,.
\eeq
This sum cannot be done explicitly. Following~\cite{kitaev} we differentiate with respect to $\Delta$ (dropping $1/N$ corrections):
\beq
\frac{1}{N}\frac{dS}{d\Delta} = \pi (2\Delta - 1) \tan(\pi \Delta)\,.
\eeq
Integrating with respect to $\Delta$ and using that the entropy at $\Delta =0$ is $N \ln 2$ gives~\cite{PhysRevB.63.134406},~\footnote{The extremal entropy~\eqref{E:SYKentropy} satisfies a nice consistency check. There is an analogue of Zamolodchikov's $c$-theorem~\cite{Zamolodchikov:1986gt} that holds in quantum mechanics: $dS/dT \geq 0$. So the extremal entropy ought to be less than the high temperature entropy $S_{free}/N = \ln 2$ at the free-field fixed point, and indeed the entropy~\eqref{E:SYKentropy} satisfies this constraint.}
\beq
\label{E:SYKentropy}
\frac{S}{N}= (1-2\Delta)\ln\left(2\cos(\pi \Delta)\right) -\frac{\text{Li}_2\left( - e^{2\pi i \Delta}\right)-\text{Li}_2\left( - e^{-2 \pi i \Delta}\right)}{2\pi i}\,.
\eeq
For $\Delta = 1/4$, this gives $S/N = G/\pi + (\ln 2)/4\approx 0.464848$, where $G$ is Catalan's constant.

With all of this in mind, there are two simple reasons why the SYK models cannot have a conventional (weakly curved, weakly coupled) gravity dual. (These reasons were also mentioned in~\cite{Maldacena:2016hyu}.)
\begin{enumerate}
	\item The entropy of the SYK models is $\mathcal{O}(N)$, so that the Newton's constant of the putative dual would be $\mathcal{O}(1/N)$. The $2N$ fermions $\psi^a$ would be dual to $2N$ degenerate, bulk fermions $\Psi^a$. However, the existence of so many light fields invalidates the saddle-point approximation: the one-loop correction to the bulk partition function from the $\Psi^a$ would be comparable to the classical saddle. 
	\item Theories with a conventional gravity dual exhibit large $N$ factorization. Consequently, given an operator $\mathcal{O}$ of dimension $\Delta$ dual to a bulk field, there are necessarily ``multi-trace'' operators e.g. $\sim \mathcal{O}(\partial^2)^n\mathcal{O}$ of dimension $2\Delta+2n+\mathcal{O}(1/N)$. Computation of the four-point function of the $\psi^a$ in the SYK models~\cite{Polchinski:2016xgd,Maldacena:2016hyu} reveals no such operators.
\end{enumerate}

These ills might be cured by gauging a large subgroup of the flavor symmetry. That is, there may yet be a \emph{gauged} SYK/AdS correspondence. This would be immensely satisfying if true. We cannot help but mention that this would be consistent with arguments that bulk locality is tied to ``large'' gauge symmetries in a field theory dual (see e.g.~\cite{ElShowk:2011ag,Mintun:2015qda}).

Even ignoring quenched disorder, gauge theory in $0+1$ dimensions is not close to the free-field fixed point owing to the integral over holonomies, and so it is not yet clear to us if the emergent conformal symmetry, large extremal entropy, and maximal chaos will persist after gauging. We are presently investigating this possibility.

\textit{Dilaton gravity.}~Two-dimensional gravity is rather different from its higher-dimensional cousins. Compactification to two dimensions generally leads to a dilaton gravity characterized by a two derivative action,
\beq
\label{E:dilatonS}
S_{\rm bulk} = \frac{1}{2\kappa^2}\int d^2x \sqrt{-g} \left( \varphi R + U[\varphi]\right) + S_{\rm matter}\,,
\eeq
where $\varphi$ is the dilaton and $U$ its potential. The equations of motion are
\begin{align}
\begin{split}
\label{E:einstein}
T_{\mu\nu} & = -D_{\mu}D_{\nu}\varphi + g_{\mu\nu}\Box \varphi - \frac{g_{\mu\nu}}{2}U  \,,
\\
\Phi & = R + U' \,,
\end{split}
\end{align}
with $T_{\mu\nu}$ and $\Phi$ the stress tensor and dilaton source,
\beq
\delta S_{\rm matter} = \frac{1}{2\kappa^2}\int d^2x \sqrt{-g} \left( T^{\mu\nu}\delta g_{\mu\nu} - \Phi \delta\varphi\right)\,.
\eeq

Dilaton gravities have AdS$_2$ vacua at roots of the dilaton potential, $U[\varphi_0]=0$ with matter fields vanishing,
\beq
\label{E:AdS2}
\varphi = \varphi_0, \,\,\, g = L^2\left( -r^2 dt^2+2dt dr\right), \,\,\, L^2 = \frac{2}{U'[\varphi_0]}.
\eeq
We take $U'[\varphi_0]=2$ hereafter. Observe that we are using infalling Eddington-Finkelstein coordinates. Holographically renormalizing in the AdS$_2$ throat~\cite{Grumiller:2007ju} shows that (i.) the dilaton is not dual to an operator, (ii.) the metric is not either, in that the dual stress tensor vanishes, and (iii.) the dual theory is invariant under a Virasoro symmetry with $c=0$~\cite{Strominger:1998yg}. The boundary theory lives at $r\to\infty$ with the metric $h = - dt^2$.

The vanishing of the boundary stress tensor is another way of stating the usual result that AdS$_2$ does not support finite-energy excitations~\cite{Maldacena:1998uz}.

Conformal symmetry is infinite-dimensional in one dimension. Any reparameterization of time $t = t(w)$ can be compensated for by a Weyl rescaling of the metric $h_{\mu\nu} \to e^{2\Omega} h_{\mu\nu}$ so as to leave the metric invariant. On the gravity side, conformal transformations correspond to diffeomorphisms which preserve the radial gauge in~\eqref{E:AdS2} and fix the boundary metric. Under the conformal transformation $t(w)$, the AdS$_2$ vacuum~\eqref{E:AdS2} becomes
\beq
\label{E:generalAdS2}
\varphi = \varphi_0, \quad g = -(r^2+2 \{ t(w),w\})dw^2+2dw dr\,,
\eeq
with $\{t(w),w\}$ the Schwarzian derivative
\beq
\{ t(w),w\} = \frac{t'''(w)}{t'(w)}- \frac{3}{2}\frac{(t''(w))^2}{(t'(w))^2}\,.
\eeq

The conformal transformation $t(w) = \tanh(\pi w T)$ has constant Schwarzian $\{ t(w),w\} = -2 \pi^2 T^2$ and maps the AdS$_2$ vacuum to an AdS$_2$ black hole
\beq
\label{E:AdSBH}
\varphi=\varphi_0\,, \qquad g = -(r^2-r_h^2)dw^2+2dw dr\,,
\eeq
with $r_h=2\pi T$ and $T$ the Hawking temperature. The thermal entropy is $S = 2\pi \varphi(r_h)/\kappa^2 = 2\pi \varphi_0/\kappa^2$.

Now consider a holographic renormalization group (RG) flow terminating in an AdS$_2$ throat. To get the basic idea, we turn off matter fields $T_{\mu\nu}=0,\Phi=0$, and try to glue the AdS$_2$ near-horizon~\eqref{E:generalAdS2} to an RG flow at large $r$. Enforcing the $rr$ component of Einstein's equations~\eqref{E:einstein}, the near-AdS$_2$ geometry is given by the perturbative solution
\begin{align}
\begin{split}
\label{E:pertSol1}
\varphi & = \varphi_0 +  \ell \left(r \varphi_1(w) + \varphi_2(w)\right) + \mathcal{O}(\ell^2 r^2)\,,
\\
g & = - \left( r^2 + 2 \{ t(w),w\} \right)dw^2 + 2 dw dr + \mathcal{O}(\ell r) \,,
\end{split}
\end{align}
where $\ell$ is a length scale satisfying $\ell r \ll 1$. The dilaton formally behaves as if it is dual to a dimension-$2$ operator, with a source $\ell \varphi_1(w)$. We work in the same spirit as~\cite{Heemskerk:2010hk,Faulkner:2010jy} and take the dual QM to ``live'' on a constant-$r$ slice at large $r\to\infty$, and fix $\varphi_1 = 1$ as a boundary condition. The $rw$ component of Einstein's equations fixes $\varphi_2 =0$, and the $ww$ component gives
\beq
\label{E:eom1}
\partial_w \big( \{ t(w),w\} \big)= 0\,.
\eeq

So, in the absence of matter, the RG flow must terminate in an AdS$_2$ black hole~\eqref{E:AdSBH}. The flow corrects the near-extremal entropy,
\beq
\label{E:nearExtremalS}
S = \frac{2\pi}{\kappa^2}\left( \varphi_0 + 2 \pi \ell T + \mathcal{O}(\ell^2T^2)\right)\,.
\eeq

Now send in matter. For simplicity, consider a small amount of infalling null dust described by a stress tensor $T_{ww}(w)\sim \ell$. The $rr$ component of Einstein's equations is unmodified, so the perturbative solution~\eqref{E:pertSol1} still holds. We again impose $\varphi_1 = 1$, and the $rw$ component fixes $\varphi_2=0$. The $ww$ component gives (to first order in $\ell$)
\beq
\label{E:eom2}
\ell\partial_w \left( \{ t(w),w\}\right) = - T_{ww}(w)\,.
\eeq
This relation is familiar: the horizon grows as matter falls in. Let us translate it into an equation in the boundary quantum mechanics. Holographically renormalizing to first order in $\ell$, we find that the boundary energy $E = - h_{\mu\nu}\langle  t^{\mu\nu}\rangle $ (with $t^{\mu\nu}$ the boundary stress tensor) is
\beq
\label{E:energy}
E = - \frac{\ell}{\kappa^2}\{ t(w),w\} \,.
\eeq
A microscopic model for the dust is a massless scalar field
\beq
S_{\rm matter} = - \frac{1}{2}\int d^2x \sqrt{-g} Z_0[\varphi](\partial\chi)^2\,,
\eeq
dual to a dimension$-1$ operator $\mathcal{O}$. The ``wavefunction renormalization'' parameter $Z_0$ satisfies $Z[\varphi_0]=1$. The infalling solutions are $\chi = \lambda(w)$ on which $T_{ww} = \kappa^2\dot{\lambda}^2$ (with $\dot{f} = \partial_w f$). The source for $\mathcal{O}$ is $\lambda(w)$, and its one-point function is $\langle \mathcal{O}\rangle = \dot{\lambda}$. Putting the pieces together,~\eqref{E:eom2} becomes 
\beq
\label{E:eom3}
\dot{E} = \dot{\lambda} \langle \mathcal{O}\rangle\,,
\eeq
which is simply the diffeomorphism Ward identity in one dimension.

We consider a general matter action in the Appendix. For a single bulk field $\chi$ dual to a dimension $\Delta$ operator $\mathcal{O}_{\Delta}$ with source $\lambda$, the Einstein's equations boil down to~\eqref{E:eom3} with the energy given by 
\beq
\label{E:generalEnergy}
E = - \frac{\ell}{\kappa^2}\{ t(w),w\} +(1-\Delta) \lambda \langle \mathcal{O}_{\Delta}\rangle\,,
\eeq
and the extension to multiple fields is obvious.

We can do better and obtain the effective action for dilaton gravity near the throat. It is
\beq
\label{E:Seff}
S_{\rm eff} =- \frac{\ell}{\kappa^2}\int dw \{t(w),w\} +W_{\rm CQM}[\lambda;t(w)]\,,
\eeq
where $W_{\rm CQM}$ is the ``generating functional'' obtained by integrating out the matter in the fixed AdS$_2$ background~\eqref{E:generalAdS2}. Equivalently, $W_{\rm CQM}$ comes from integrating out matter in the pure AdS$_2$ geometry~\eqref{E:AdS2}, followed by a conformal transformation $t(w)$. Here $t(w)$ is the fundamental field and its Euler-Lagrange equation is~\eqref{E:eom3}.

\textit{Hydrodynamics.}~This result evokes the fluid/gravity correspondence~\cite{Bhattacharyya:2008jc}, in that we have rewritten the gravitational dynamics as the (non-)conservation of energy in the boundary quantum mechanics with a ``constitutive relation''~\eqref{E:generalEnergy} for the energy. Unlike the fluid/gravity correspondence, this rewriting does not rely on a gradient expansion or even a black hole to start with.

Let us take this connection to hydrodynamics seriously.

Haehl, Loganayagam, and Rangamani (HLR) have classified~\cite{Haehl:2015pja} the most general hydrodynamics consistent with the second Law of thermodynamics, building upon earlier results in hydrostatic equilibrium~\cite{Banerjee:2012iz,Jensen:2012jh}. HLR also obtained Schwinger-Keldysh effective actions~\cite{Haehl:2015uoc} for hydrodynamics (see also~\cite{Crossley:2015evo}). A subset of allowed transport (what they dub class $\sf L$, for Lagrangian) admits an ordinary action via a sigma model, where the fundamental fields are maps from a ``reference manifold'' to the physical spacetime~\cite{Haehl:2015pja}.

The effective action~\eqref{E:Seff} for dilaton gravity is just such a class $\sf L$ action. Recall that $t(w)$ is the conformal transformation from the AdS$_2$ vacuum to the state of the system. It is useful to redefine $t(w) = \tanh (\pi \sigma(w)/\beta)$ so that $\sigma(w)$ is the fundamental field, which represents a conformal transformation starting from the thermal state with temperature $1/\beta$. In terms of $\sigma(w)$ and after an integration by parts, the effective action~\eqref{E:Seff} becomes
\beq
S_{\rm eff} =  \frac{\ell}{2\kappa^2}\int dw \left(\frac{\sigma''(w)^2}{\sigma'(w)^2} +\frac{4\pi^2}{\beta^2}\sigma'(w)^2\right) + W_{\rm CQM}\,.
\eeq
We take $w$ to be the coordinate on the physical spacetime $\mathcal{M}$, and $\sigma$ parameterizes the ``reference manifold'' $\sf M$. The metric on $\sf M$ is ${\sf h}=  - w'(\sigma)^2 d\sigma^2$, and on $\sf M$ we define the fixed vector field $\bm \beta^\sigma =\beta $. From this data we define a time-dependent temperature and velocity
\beq
 {\sf T} = \frac{1}{\sqrt{-{\sf h}_{ab} {\bm \beta^a}{\bm \beta^b}}} \,, \qquad {\sf u}^a = \frac{{\bm \beta}^a}{\sqrt{-{\sf h}_{bc}{\bm \beta}^b{\bm \beta}^c}}\,,
 \eeq
and $\dot{f} = {\sf u}^a \partial_a f$. Then
\beq
\label{E:Shydro}
S_{\rm eff} = \int d\sigma \sqrt{-\sf h}\left\{\sf P(\sf T) + \frac{\ell}{2\kappa^2} \frac{\sf\dot{T}^2}{\sf T^2} \right\} + W_{\rm CQM}\,,
\eeq
where $\sf P(T)$ is the pressure
\beq
{\sf P(\sf T)} =-E_0+ \frac{2\pi}{\kappa^2}\left(\varphi_0 {\sf T} + \pi \ell {\sf T}^2 \right)\,,
\eeq
and $E_0$ is the ground state energy. (Strictly speaking, neither the ground state energy nor linear term was present in~\eqref{E:Seff}; but neither affects the equation of motion and so we lose nothing by adding them.) Reparameterization invariance guarantees that the equation of motion for $w(\sigma)$, keeping $h$ and $\bm \beta^a$ fixed, is precisely~\eqref{E:eom3}.

A few comments are in order.
\begin{enumerate}
\item The hydrodynamic action also computes the low-temperature free energy~\footnote{Maldacena and Stanford~\cite{Maldacena:2016hyu} have also obtained a Schwarzian low-energy effective action for the SYK models. They were the first to observe that said action also computes the near-extremal thermodynamics.}. Wick-rotating to Euclidean signature, the action evaluated on the solution $w(\sigma) = \sigma$ (so that ${\sf T} = T$) gives
\begin{equation*}
\ln \mathcal{Z}_E = i S_E = - \beta E_0 + \frac{2\pi}{\kappa^2}\left( \varphi_0 + \pi \ell T + \mathcal{O}(\ell^2 T^2)\right)\,.
\end{equation*}
\item The $\dot{\sf T}^2$ and ${\sf T}^2$ terms in the hydrodynamic action are linked: they arise from the Schwarzian action~\eqref{E:Seff} after conformally transforming from the vacuum. In this way the low-temperature correction to the entropy (equivalently a low-energy correction to the density of states) determines the dynamics. In principle there are higher derivative corrections to the $\mathcal{O}(\ell)$ hydrodynamic action~\eqref{E:Shydro} e.g. $\frac{\ell}{\kappa^2}\frac{\ddot{{\sf T}}^2}{{\sf T}^4}$. However, as far as we can tell, all such terms are forbidden by demanding regularity in the vacuum (as long as $\sigma'(w)>0$). In this sense the $\mathcal{O}(\ell)$ hydrodynamic action seems to be unique.
\item At $\mathcal{O}(\ell^2)$ however we expect there to be additional terms in $S_{\rm eff}$, like $\ell^2 {\sf T}^3$.
\item It would be interesting to go beyond the classical limit and compute quantum corrections to the free energy, correlators, \&c, arising from the hydrodynamic mode $w(\sigma)$.
\end{enumerate}

\textit{Four-point functions and chaos.}~One of the basic entries in the holographic dictionary is the computation of CFT correlation functions via Witten diagrams in AdS. For dilaton gravity near AdS$_2$, the computation of two and three-point functions of boundary operators is straightforward, and the result is the usual one dictated by conformal invariance. The four-point function is much richer. It has two parts. The first is a conformally invariant contribution involving a sum over conformal blocks, dual to tree-level contact and exchange Witten diagrams. The second breaks conformal invariance, dominates the first, and is due to the hydrodynamics~\eqref{E:Shydro}. What happens is this: quadratic fluctuations of the source $\lambda$ for the operator $\mathcal{O}$ inject energy: they source the ``Goldstone mode'' $w(\sigma)$. Plugging the fluctuation $\delta w(\sigma)\sim \lambda^2$ back into the matter action $W_{\rm CQM}$ leads to an $\mathcal{O}(\lambda^4)$ contribution to the on-shell action. 

We stress that this ``hydrodynamic backreaction'' and the concomitant conformal symmetry breaking was anticipated by Almheiri and Polchinski~\cite{Almheiri:2014cka}, who studied a soluble toy model of two-dimensional holography.

We illustrate the importance of this hydrodynamic contribution by computing the Lyapunov exponent at tree-level in the effective description. Consider an out-of-time-ordered, connected, thermal four-point function~\cite{1969JETP...28.1200L,Shenker:2013pqa} of two operators $W$ and $V$,
\beq
\label{E:outOfTime}
F(w) \equiv \langle W(w)V(0)W(w)V(0)\rangle_{\beta} \,.
\eeq
The Lyapunov exponent $\lambda_L$ characterizes the growth of $F(w)\sim  e^{\lambda_L w}$. We obtain $F(w)$ from the Euclidean vacuum four-point function by the same method as in~\cite{Roberts:2014ifa,Polchinski:2016xgd}.

We begin on the Euclidean line $\bar{\tau}$ and turn on a source $\lambda$ for $\mathcal{O}_{\Delta}$, normalized as $\langle \mathcal{O}_{\Delta}(\tau)\mathcal{O}_{\Delta}(0)\rangle = 1/|\bar{\tau}|^{2\Delta}$. Turning on $\lambda$ sources a conformal transformation $\bar{\tau}(\tau)$. The conformally transformed $W_{\rm CQM}$ is
\beq
W_{\rm CQM} = \frac{1}{2}\int \frac{d\tau_1 d\tau_2 \big( \bar{\tau}'(\tau_1)\bar{\tau}'(\tau_2)\big)^{\Delta}}{|\bar{\tau}(\tau_1) - \bar{\tau}(\tau_2)|^{2\Delta}} \lambda(\tau_1)\lambda(\tau_2) + \mathcal{O}(\lambda^3),
\eeq
With $\bar{\tau}(\tau) = \tau + \varepsilon(\tau)$, the equation of motion~\eqref{E:eom3} gives
\begin{align}
\varepsilon(\tau) = &\frac{\kappa^2\Delta }{12\ell} \int \frac{d\tau_1 d\tau_2}{|\tau_1-\tau_2|^{2\Delta}}|\tau-\tau_1|^3
\\
\nonumber
&  \times \left\{ \frac{3}{\tau-\tau_1} +\frac{2}{\tau_1-\tau_2} \right\}\lambda(\tau_1)\lambda(\tau_2) + \mathcal{O}(\lambda^3)\,.
\end{align}
Feeding this back into $S_{\rm eff}$ leads to an $\mathcal{O}(\lambda^4)$ term, both from the Schwarzian and from the variation of $W_{\rm CQM}$:
\beq
\label{E:deltaSeff}
\delta_{\varepsilon} S_{\rm eff} =\int d\tau \left( \frac{\ell}{2\kappa^2}\ddot{\varepsilon}^2 -\Delta \dot{\varepsilon} \lambda\langle \mathcal{O}_{\Delta}\rangle - \varepsilon\lambda\langle \dot{\mathcal{O}}_{\Delta}\rangle\right),
\eeq
and so the hydrodynamic four-point function. Observe that for $1/\kappa^2 \sim N$, this contribution is $1/(N \ell)$, which is $\mathcal{O}(1/N)$ and breaks conformal invariance as advertised.

From this, we find the Euclidean, hydrodynamic four-point function for two different operators $W$ and $V$
\begin{widetext}
\beq
\label{E:4pt}
\frac{\langle W(\tau_1)W(\tau_2)V(\tau_3)V(\tau_4) \rangle }{\langle W(\tau_1)W(\tau_2)\rangle \langle V(\tau_3)V(\tau_4)\rangle}= \frac{\kappa^2}{\ell}\Delta_W\Delta_V\left\{ |\tau_{13}|^3\left( \frac{2}{3\tau_{12}\tau_{34}}+\frac{1}{\tau_{12}\tau_{13}}+\frac{1}{\tau_{13}\tau_{34}}\right)-|\tau_{13}| +(\text{permutations})\right\}\,,
\eeq
\end{widetext}
with $\tau_{ij}=\tau_i-\tau_j$. We conformally map to the thermal state $\tau = \tanh( \pi w/\beta)$ and take the ``second sheet'' analytic continuation
\beq
w_1 = w+2i \epsilon\,, \,\, w_2 = w -i\epsilon\,, \,\, w_3 = i \epsilon\,, \,\,w_4 = -i \epsilon\,.
\eeq
The terms in brackets cancel against their permutations, and the remaining term goes as $|\tau_{13}|$, so that the four-point function grows as $\tau_1\sim \tanh(\pi w T)\sim \exp(2\pi w T)$. We thereby extract
\beq
\lambda_L = 2\pi T\,.
\eeq

\textit{Conclusions.}~We have found two main results. First, the SYK models do not have a conventional gravity dual, although perhaps there is a gauged SYK/AdS correspondence. Second, with the prospect of such a correspondence in mind, we unraveled various thorny issues in AdS$_2$ holography. Our central result was to rewrite the gravitational dynamics near an AdS$_2$ throat in terms of an effective quantum mechanical action~\eqref{E:Seff}. This action was that of a novel hydrodynamics~\eqref{E:Shydro} coupled to CQM correlators. Unlike ordinary hydrodynamics it describes the dynamics all the way down to extremality. 

This hydrodynamics is intimately tied up with diffeomorphism invariance. The sole hydrodynamic mode ensures that the diffeomorphism Ward identity is satisfied in the infrared. The hydrodynamic description plays a similar role in two-dimensional holography as the Virasoro identity block with $c\gg 1$ does in AdS$_3$: both approximate the leading contribution to four-point functions in their respective field theory duals. 

Very recently, Maldacena and Stanford~\cite{Maldacena:2016hyu} have obtained the same Schwarzian effective action~\eqref{E:Seff} from the large $N$ solution of the SYK models. The emergence of the same description in two rather different systems raises the question of its universality.

In the main text we suggested that the leading low-energy part of the hydrodynamic action was unique, in the sense that the coefficients of the $\dot{ {\sf T}}^2$ and ${\sf T}^2$ terms were linked as in~\eqref{E:Shydro} and that there are no gradient corrections. 

If this is the case, then it seems reasonable that the hydrodynamic description universally describes (diffeomorphism-invariant) large $N$ systems with an emergent conformal invariance, and consequently any such system will be maximally chaotic.


\textit{Acknowledgements.} This work grew out of a collaboration with C.~P.~Herzog, to whom I owe many thanks. It is also a pleasure to thank A.~O'Bannon, L.~Rastelli, M.~Rozali, S.~Sachdev, and S.~Shenker for enlightening discussions.


\section{Appendix}


\subsection{A. Conformal symmetry in one dimension}


Let us briefly recap some basic features of conformal invariance in one dimension. It is helpful to keep in mind that, in interacting systems like the SYK models, this conformal invariance is only an approximate symmetry of certain large $N$ theories.

Consider coupling a CQM to an external metric $h_{\mu\nu}$ and a source $\lambda$ for a scalar operator $\mathcal{O}_{\Delta}$ of dimension $\Delta$. Define the generating functional of connected correlation functions,
\beq
W = - i \ln \mathcal{Z}\,.
\eeq
The connected one-point functions of the stress tensor $t^{\mu\nu}$ and $\mathcal{O}_{\Delta}$ are defined by functional variation,
\beq
\label{E:deltaW}
\delta W = \int dt \sqrt{-h} \left( \frac{1}{2}\langle t^{\mu\nu} \rangle \delta h_{\mu\nu} - \langle \mathcal{O}_{\Delta}\rangle \delta \lambda\right)\,.
\eeq
By assumption, $W$ is invariant under infinitesimal reparameterizations, under which $h_{\mu\nu}$ and $\lambda$ vary as
\beq
\delta_{\xi} h_{\mu\nu} =D_{\mu}\xi_{\nu}+D_{\nu}\xi_{\mu}\,, \qquad \delta_{\xi}\lambda = \xi^{\mu}D_{\mu}\lambda\,,
\eeq
with $D_{\mu}$ the covariant derivative. Plugging these variations into~\eqref{E:deltaW} and demanding $\delta_{\xi}W=0$ leads to the diffeomorphism Ward identity
\beq
D_{\nu}\langle t^{\mu\nu} \rangle = - (D^{\mu}\lambda)\langle \mathcal{O}_{\Delta}\rangle\,.
\eeq
In one dimension with $h = - dt^2$, this becomes equation~\eqref{E:eom3} from the main text,
\beq
\label{E:diffWard}
\dot{E} =  \dot{\lambda} \langle \mathcal{O}_{\Delta}\rangle\,.
\eeq

On the other hand, we ought to couple the CQM to the metric in a Weyl-invariant way. Then $W$ is invariant under an infinitesimal Weyl rescaling under which the metric and source $\lambda$ vary as
\beq
\delta_{\omega} h_{\mu\nu} = 2 \omega h_{\mu\nu}\,, \qquad \delta_{\omega}\lambda = (\Delta - 1)\omega \lambda\,,
\eeq
and $\delta_{\omega}W=0$ gives the Weyl Ward identity (in terms of $E = - h_{\mu\nu}\langle t^{\mu\nu}$)
\beq
\label{E:weylWard}
E = (1-\Delta)\lambda \langle \mathcal{O}_{\Delta}\rangle\,.
\eeq

One way of stating Polcinski's paradox~\cite{Jensen:2011su} is that the diffeomorphism Ward identity~\eqref{E:diffWard} is not compatible with the Weyl Ward identity~\eqref{E:weylWard} in one dimension. Actually there is a loophole: the two are compatible if the correlation functions of $\mathcal{O}$ are topological, as they are for operators in the theory of $2N$ free Majorana fermions.

So there is a conflict between reparameterization invariance and conformal symmetry for interacting systems in one dimension.

Before going on, we should reiterate what is in some sense the main point of this Letter, namely how gravity solves this paradox near AdS$_2$. The boundary dual is not a CQM on its own, but instead a particular hydrodynamics coupled to the correlation functions of a CQM. 

Those conformal correlators are strongly constrained by the conformal symmetry, as we now discuss. In any dimension, conformal transformations are the combination of a coordinate transformation $x^{\mu}=x^{\mu}(y^{\nu})$ and Weyl rescaling $h_{\mu\nu} \to e^{2\Omega}h_{\mu\nu}$ which leave the metric invariant. In one dimension with the flat metric $h=-dt^2$, any coordinate transformation gives a conformal transformation: the combined action of
\beq
t = t(w)\, \qquad \Omega = - \ln t'(w)\,,
\eeq 
sends the metric to itself, $-dt^2 \to -dw^2$. Wick-rotating to Euclidean signature and compactifying Euclidean time, the modes of $t(w)$ generate a Virasoro algebra with $c=0$. 

There are two ways to think about the vanishing central charge. The first is simply that the stress tensor vanishes by the Weyl Ward identity. The second is that there is no Weyl anomaly possible in one dimension.

As is familiar from two-dimensional CFT, the conformal transformations which are regular everywhere generate the global conformal group $SL(2;\mathbb{R})$. There is a single special conformal generator $K$, and in the usual way one defines primary operators as those annihilated by $K$. The vacuum two-point function of a primary operator $\mathcal{O}_{\Delta}$ of dimension $\Delta$ is, in Euclidean signature
\beq
\langle \mathcal{O}_{\Delta}(\tau)\mathcal{O}_{\Delta}(0)\rangle = \frac{1}{|\tau|^{2\Delta}}\,,
\eeq
and similarly for three-point functions of $\mathcal{O}_{\Delta}$.

We conclude this Subsection with a few comments on thermodynamics. Consider the thermal partition function of a CQM
\beq
\mathcal{Z}_E = \text{tr}\left( e^{-\beta H}\right)\,,
\eeq
which as usual is the partition function of the Euclidean theory on a circle of size $\beta$ with thermal boundary conditions. There is only one local counterterm which can be used to redefine the theory,
\beq
\label{E:CT}
\ln \mathcal{Z}_E \to \ln\mathcal{Z}_E + m\int d\tau \sqrt{h_E}\,,
\eeq
where $\tau$ is Euclidean time, $h_E$ the Euclidean metric, and $m$ is some mass scale. This counterterm is not scale-invariant. Thus, in a CQM, the thermal partition function $\mathcal{Z}_{CQM}$ is an unambiguous observable. This should not be a surprise; in odd dimension, the logarithm of the partition function of a CFT on a Euclidean sphere -- the ``sphere free energy'' -- is a useful and unambiguous (up to a quantized imaginary ambiguity which may exist in $d=4k-1$ dimensions) CFT observable. In one dimension, this partition function is the extremal entropy,
\beq
S_0 = \frac{\partial (T\ln \mathcal{Z}_{CQM})}{\partial T} = \ln \mathcal{Z}_{CQM}\,.
\eeq 

Now consider a non-conformal QM which realizes an emergent conformal invariance at low energies. Suppose that the low-energy description is a(n approximate) CQM with extremal entropy $S_0$ deformed by a dimension $\Delta$ operator. The low-temperature partition function is
\beq
\ln \mathcal{Z}_E = - \beta E_0 + S_0 + P_1 T^{\Delta-1} + \hdots\,,
\eeq
where $E_0$ is the ground state energy. Observe that $E_0$ is unphysical: it is redefined by the counterterm~\eqref{E:CT}. This partition function leads to a low-temperature entropy
\beq
S = S_0 +\Delta  P_1 T^{\Delta-1} + \hdots\,.
\eeq
Comparing this with the entropy~\eqref{E:nearExtremalS} of near-extremal black holes in dilaton gravity, we see that the black hole entropy is that of a CQM deformed by a $\Delta = 2$ operator.


\subsection{B. Holographic renormalization}


In this Appendix we fill in various details on the near-AdS$_2$ solutions described in the main text.

We begin with the perturbative solution near the endpoint of a holographic RG flow, with a small amount of infalling dust $T_{ww}(w)$. The full perturbative solution is
\begin{align}
\nonumber
\varphi & = \varphi_0 + \ell r + \mathcal{O}(\ell^2 r^2)\,,
\\
\nonumber
g & = - \left(r^2 + 2 \{ t(w),w\} +  \frac{\ell r^3}{6}U''[\varphi_0] + \mathcal{O}(\ell^2r^2)\right)dw^2 
\\
\label{E:perturbativeSolution}
& \qquad + 2dw dr\,,
\end{align}
with
\beq
\label{E:bulkWard}
\ell\partial_w \{ f(w),w\}  = -T_{ww}(w)\,.
\eeq

We account for the dust with a massless scalar field $\chi$
\begin{equation*}
S_{\rm matter} = - \frac{1}{2}\int d^2x \sqrt{-g} Z_0[\varphi] (\partial\chi)^2\,,
\end{equation*}
with $Z[\varphi_0]=1$. The infalling solutions are $\chi(w) = \lambda(w)$, and we take the perturbative limit $\lambda^2 \sim\ell$. To first order in $\ell$, the stress tensor evaluated on this solution is $T_{ww} = \kappa^2\dot{\lambda}^2$ and the dilaton source vanishes $\Phi = 0$.

We proceed to holographically renormalize the dilaton gravity~\eqref{E:dilatonS} on backgrounds of the form~\eqref{E:perturbativeSolution}. To proceed we define the dual theory on a constant-$r$ slice at asymptotically large $r$, subject to the constraint that $\ell r$ is still asymptotically small. In physical terms, we are taking the almost zero energy limit. In practice we drop the corrections to the background~\eqref{E:perturbativeSolution} with two or more powers of $\ell$ and then proceed in the usual way.

The bulk action is divergent as one integrates to large $r$, so we introduce a cutoff slice at $r=\Lambda$, add covariant boundary terms on the cutoff slice to eliminate divergences, and then remove the cutoff by sending $\Lambda\to\infty$.

The authors of~\cite{Grumiller:2007ju} have studied the problem of holographic renormalization in a general dilaton gravity. For the case at hand, the renormalized action is simply
\begin{align}
S_{ren} &= \lim_{\Lambda\to\infty} \frac{1}{2\kappa^2}\left\{ \int d^2x \sqrt{-g} \left( \varphi R + U\right)\right.
\\
\nonumber
&  \qquad\qquad + \left. \int dt \sqrt{-\gamma} \left( 2\varphi K - U \right) \right\} + S_{\rm matter} + \mathcal{O}(\ell^2)\,,
\end{align}
where $\gamma$ is the induced metric on the cutoff slice and $K$ the trace of its extrinsic curvature.

The variation of with on-shell action with respect to the metric and scalar is
\begin{align}
\begin{split}
\label{E:deltaSren}
\delta S_{ren} =& \lim_{\Lambda\to\infty} \int dt \sqrt{-\gamma}\left\{ - (n^{\mu}\partial_{\mu}\chi)\delta\chi\right.
\\
& \qquad \quad \left. +\frac{1}{2\kappa^2} \left( n^{\rho}\partial_{\rho}\varphi - \frac{U}{2}\right)\gamma^{\mu\nu}\delta g_{\mu\nu}\right\}\,,
\end{split}
\end{align}
with $n^{\mu}$ the outward pointing normal vector to the cutoff slice. Both $n^{\mu}\partial_{\mu}\varphi$ and $U$ are $\mathcal{O}(\ell)$ for the perturbative solution~\eqref{E:perturbativeSolution}, so we require the on-shell variation of the $g_{\mu\nu}$ in response to a variation of the boundary metric to $\mathcal{O}(\ell^0)$, which is simply $\delta g_{\mu\nu} = r^2 \delta h_{\mu\nu} + \mathcal{O}(r^0,\ell)$. Plugging this variation into $\delta S_{ren}$ and evaluating it on the solution~\eqref{E:perturbativeSolution} gives the boundary stress tensor
\beq
\langle t^{\mu\nu}\rangle = \frac{2}{\sqrt{-h}}\frac{\delta S_{ren}}{\delta h_{\mu\nu}} = \frac{\ell h^{\mu\nu}}{\kappa^2}\{t(w),w\}\,,
\eeq
so that the energy is $E = - h_{\mu\nu} \langle t^{\mu\nu}\rangle = - (\ell/\kappa^2) \{ t(w),w\}$, which derives the result~\eqref{E:energy} in the main text. We also obtain the expectation value of the dimension$-1$ operator $\mathcal{O}$ dual to $\chi$,
\beq
\langle \mathcal{O}\rangle = - \frac{1}{\sqrt{-h}}\frac{\delta S_{ren}}{\delta \lambda} = \dot{\lambda}\,,
\eeq
so that~\eqref{E:bulkWard} becomes
\begin{equation*}
\dot{E} = \dot{\lambda}\langle \mathcal{O}\rangle\,.
\end{equation*}

Now we consider the perturbative solution in the presence of massive matter,
\beq
S_{\rm matter} = - \frac{1}{2}\int d^2x \sqrt{-g} \left( Z_0[\varphi] (\partial\chi)^2 + Z_1[\varphi]m^2 \chi^2\right)\,,
\eeq
with $Z_0[\varphi_0]=Z_1[\varphi_0]=1$. The field $\chi$ is now dual to an operator $\mathcal{O}_{\Delta}$ of dimension $\Delta(\Delta - 1)=m^2$. We take $\chi$ to be a free field, but it is easy to allow for self-interactions.

As above we take the perturbative limit with $\chi \ll \ell$. In this limit,
\begin{align}
\begin{split}
T_{\mu\nu} & = \kappa^2 \left( \partial_{\mu}\chi\partial_{\nu}\chi - \frac{ (\partial\chi)^2 +m^2}{2}g_{\mu\nu}\right)\,,
\\
\Phi & = \kappa^2 \left( Z_0'[\varphi_0](\partial\phi)^2 + Z_1'[\varphi_0]m^2 \chi^2\right)\,.
\end{split}
\end{align}
The dilaton and metric are perturbed as
\begin{align}
\nonumber
\varphi &= \varphi_0 + \ell  \psi(w,r) + \mathcal{O}(\ell^2 r^2)\,,
\\
\nonumber
g & = -\Big(r^2 + 2 \{ t(w),w\} + \ell  f(w,r)+ \mathcal{O}(\ell^2 r^2)\Big)dw^2 
\\
\label{E:perturbativeSolution2}
& \qquad + 2 dw dr \,.
\end{align}
To leading order in $\ell$, the solution for $\chi$ is given by the most general solution of
\beq
(\Box - m^2) \chi = 0\,,
\eeq
on the AdS$_2$ background~\eqref{E:perturbativeSolution2}, setting all of the $\mathcal{O}(\ell)$ corrections to vanish. We take $\Delta$ to be general but not half-integer, and pick the standard quantization for $\chi$ so that $\Delta>1/2$. Then near the boundary $r\to\infty$ $\chi$ is
\beq
\label{E:bdyChi}
\chi = r^{\Delta - 1}  \sum_{n=0}\frac{a_n(w)}{r^n}+ r^{-\Delta} \sum_{m=0}\frac{b_m(w)}{r^m}\,,
\eeq
where the $a_{n}$ with $n>0$ are determined by $a_0$ and similarly for the $b_n$, for example
\beq
a_1 = \dot{a}_0\,, \qquad b_1 = \dot{b}_0\,.
\eeq
We identify the source for the dual operator as
\beq
\lambda = \lim_{r\to\infty} r^{1-\Delta} \chi = a_0(t)\,.
\eeq

The stress tensor and dilaton source have three distinct sets of terms. Near the boundary, they are of the schematic form
\beq
T_{\mu\nu} = r^{2\Delta}\Sigma^{a}_{\mu\nu}(r,a^2) + \Sigma_{\mu\nu}(r,ab) + r^{-2\Delta} \Sigma^b_{\mu\nu}(r,b^2)\,,
\eeq
where $\Sigma^a_{\mu\nu}(r,a^2)$ is some power series in $1/r$ with coefficients built out of two powers of the $a_n$ and their derivatives, and similarly for the other two series. Because $\Delta\cnot \in \mathbb{Z}/2$ by assumption, these three series never mix.

Solving the $rr$ and $rw$ components of Einstein's equations, we obtain the correction to the dilaton to be
\begin{align}
\label{E:psi}
&\psi = r + \frac{\kappa^2}{\ell}r^{2(\Delta -1)}\left( \frac{(\Delta-1)a_0^2}{2(3-2\Delta)} + \mathcal{O}(r^{-1})\right)
\\
\nonumber
&+ \frac{\kappa^2}{\ell r}\left(\Delta(\Delta-1) a_0 b_0 +  \frac{(\Delta^2-1)\dot{(a_0b_0)}-\Delta \dot{a}_0b_0}{3r} + \mathcal{O}(r^{-2})\right)
\\
\nonumber
& \qquad - \frac{\kappa^2}{\ell^2}r^{-2\Delta}\left(  \frac{\Delta b_0^2}{2(2\Delta+1)} + \mathcal{O}(r^{-1})\right)\,,
\end{align}
and, while the correction to the metric is calculable, we do not require it for what follows. The $ww$ component imposes exactly one condition,
\beq
\label{E:generalHydro}
\frac{\ell}{\kappa^2} \partial_w \left( \{t(w),w\}\right) =(2\Delta-1) \left( \Delta \dot{\left(a_0 b_0\right)}  -a_0\dot{b}_0\right)\,.
\eeq

As for a massless $\chi$, we have succeeded in rewriting the dilaton equations of motion in terms of an equation~\eqref{E:generalHydro} which only involves the boundary time. Let us rewrite it in terms of boundary variables.

There are additional $\mathcal{O}(\chi^2)$ boundary counterterms required to renormalize the action. The leading divergence is removed by the counterterm,
\beq
\label{E:leadingCT}
S_{\rm CT} = \frac{\Delta -1}{2}\int dt \sqrt{-\gamma}\,\chi^2 + \mathcal{O}(\partial^2 \chi^2)\,,
\eeq
and there are subleading counterterms with at least two boundary derivatives which remove subleading divergences. Varying the on-shell action with respect to the boundary metric $h$ and source $\lambda$ gives
\begin{align}
\begin{split}
\label{E:constitutiveGeneral}
\langle \mathcal{O}_{\Delta}\rangle &= (1- 2\Delta) b_0(t)\,,
\\
E &  = - \frac{\ell}{\kappa^2}\{ t(w),w\} + (2\Delta-1)(\Delta-1)a_0b_0\,,
\\
& = - \frac{\ell}{\kappa^2}\{t(w),w\} + (1-\Delta)\lambda \langle \mathcal{O}_{\Delta}\rangle\,.
\end{split}
\end{align}
(The additional $a_0b_0$ terms in the energy come from (i.) the $\mathcal{O}(1/r)$ correction to the dilaton~\eqref{E:psi}, inserted into the metric variation of $S_{ren}$ in~\eqref{E:deltaSren}, and (ii.) the metric variation of the leading counterterm~\eqref{E:leadingCT}.)
In terms of $E$, $\langle \mathcal{O}_{\Delta}\rangle$, and $\lambda=a_0$, the equation of motion~\eqref{E:generalHydro} is simply
\beq
\label{E:diffAgain}
\dot{E} = \dot{\lambda}\langle \mathcal{O}_{\Delta}\rangle\,.
\eeq

It is straightforward to allow for half-integer $\Delta$. When $\Delta\in \mathbb{Z}/2$ there are logarithmic terms in the near-boundary solution for bulk fields as well as logarithmic boundary counterterms. It is similarly straightforward to consider self-interacting matter, e.g. $\chi^4$ theory. The final result is the same: the Einstein's equations can be rewritten as the diffeomorphism Ward identity in the dual quantum mechanics for \emph{any} value of $\Delta$.

As in the main text, we have shown that the gravitational dynamics near AdS$_2$ can be rewritten as the diffeomorphism Ward identity~\eqref{E:diffAgain} with a ``constitutive relation'' for the energy given by~\eqref{E:constitutiveGeneral}. This equation of motion follows from the same effective action~\eqref{E:Seff} presented in the main text.

We see that the hydrodynamic effective action~\eqref{E:Seff} (equivalently,~\eqref{E:Shydro}) encodes the gravitational dynamics for dilaton gravity coupled to general scalar matter.


\subsection{C. Electric AdS$_2$}

In this main text, we considered dilaton gravities with AdS$_2$ vacua at the roots of the dilaton potential. There is another way to realize AdS$_2$ solutions: the AdS$_2$ may be supported by an electric field. In this Appendix we work out various aspects of these ``electric'' AdS$_2$ vacua of dilaton gravities with a Maxwell field, Einstein-Maxwell-Dilaton gravity.

Our motivation for studying this problem is somewhat indirect. Hartman and Strominger~\cite{Hartman:2008dq} have claimed that the dual to gravity on electric AdS$_2$ vacua is invariant under a Virasoro algebra with \emph{nonzero} central charge, a claim which was supported by way of holographic renormalization in~\cite{Castro:2008ms} (although see also~\cite{Castro:2014ima}). 

In the setting of two-dimensional CFT, Roberts and Stanford~\cite{Roberts:2014ifa} have shown that large central charge,  a large gap for higher-spin Virasoro primary operators, and a ``low'' density of light states is enough to guarantee a maximal Lyapunov exponent $\lambda_L = 2\pi T$. Their argument would go through more or less unaltered for a CQM invariant under a Virasoro symmetry with large central charge, assuming that such a thing existed.

Thus we revisit electric AdS$_2$ vacua and the claim of a Virasoro algebra with $c \neq 0$. We will find that the central charge of the Virasoro symmetry vanishes.

Consider the most general Einstein-Maxwell-Dilaton gravity,
\beq
S_{\rm bulk} = \frac{1}{2\kappa^2}\int d^2x \sqrt{-g} \left( \varphi R + U[\varphi]  - \frac{W[\varphi]}{4}F^2\right)\,.
\eeq  
The authors of~\cite{Castro:2008ms} consider the case
\beq
U=8\varphi\,, \qquad W = 1\,.
\eeq
The equations of motion are now
\begin{align}
\nonumber
0 & = D_{\mu}D_{\nu}\varphi - g_{\mu\nu}\Box \varphi +\frac{g_{\mu\nu}}{2} \left(U - \frac{W}{4}F^2\right) + \frac{W}{2}F_{\mu\rho}F_{\nu}{}^{\rho}\,,
\\
\label{E:einstein2}
0 & = D_{\nu}\left( WF^{\mu\nu}\right)\,,
\\
\nonumber
0 & = R + U'  - \frac{W'}{4}F^2\,.
\end{align}
They admit ``electric'' AdS$_2$ solutions with constant dilaton and constant electric field,
\beq
\varphi = \varphi_0\,, \qquad F_{\mu\nu} = E \varepsilon_{\mu\nu}\,, \qquad R = - \frac{2}{L^2}\,,
\eeq
provided the dilaton $\varphi_0$, electric field $E$, and AdS radius $L$ satisfy the two relations
\beq
\label{E:familyOfAdS2}
U[\varphi_0] = \frac{W[\varphi_0]}{2}E^2\,, \quad \frac{2}{L^2} = U'[\varphi_0] + \frac{W'[\varphi_0]E^2}{2}\,.
\eeq
So, adjusting the electric field simultaneously adjusts the dilaton and AdS radius. Observe that, for smooth potentials $U$ and $W$, there is generally a one-parameter family of AdS$_2$ solutions controlled by the strength of the electric field. The most general such solution is, in a radial gauge,
\begin{align}
\begin{split}
\label{E:electricAdS2}
\varphi & = \varphi_0\,,
\\
A & =  - EL^2 r \left( f_1(t) - \frac{f_2(t)}{r^2}\right)dt + a(t) dt\,,
\\
g & = L^2 \left( - r^2 \left( f_1(t) + \frac{f_2(t)}{r^2}\right)^2 dt^2 + \frac{dr^2}{r^2}\right)\,.
\end{split}
\end{align}
Using the defining function $1/(r^2L^2)$ the conformal boundary $r\to\infty$ is endowed with a metric
\beq
h = \lim_{r\to\infty} \frac{\gamma}{r^2L^2} = - f_1(t)dt^2\,,
\eeq
where $\gamma$ is the induced metric on a constant-$r$ slice.

Now let us holographically renormalize the bulk theory. Consider the following definition of the renormalized theory:
\begin{align}
\begin{split}
\label{E:S1}
S_{1} &= \lim_{\Lambda\to\infty}  \frac{1}{2\kappa^2}\left\{\int d^2x \sqrt{-g} \Big( \varphi R + U - \frac{W}{4}F^2\Big) \right.
\\
& \qquad  \quad\quad  +\left. \int dt \sqrt{-\gamma} \Big( 2\varphi K - LU + \frac{W}{2L}A^2\Big)\right\},
\end{split}
\end{align}
where $A^2 = \gamma^{\mu\nu}A_{\mu}A_{\nu}$. This matches the renormalization scheme of~\cite{Castro:2008ms} for the particular dilaton theory they study. It is easy to verify that this renormalized action is finite on-shell. Under a variation of the metric and gauge field, keeping the dilaton fixed and using that $\partial_{\mu}\varphi = 0$ for the solutions at hand, $S_1$ varies as
\begin{align}
\nonumber
\delta S_1 =& \lim_{\Lambda\to\infty} \frac{1}{2\kappa^2}\int dt \sqrt{-\gamma} \left\{ -\frac{L}{2}\left(   U+\frac{W}{2L^2}A^2\right)\gamma^{\mu\nu} \delta g_{\mu\nu} \right.
\\
\label{E:deltaS1}
& \qquad \qquad \left. + W \left( F^{\mu\nu} n_{\nu} + \frac{\gamma^{\mu\nu}}{L}A_{\nu}\right) \delta A_{\mu}\right\}\,.
\end{align}
On-shell, a variation of the boundary metric $h = - f_1(t)^2 dt^2$ induces a variation of both the bulk metric and gauge field~\eqref{E:electricAdS2} according to $\delta f_1 = - \frac{\delta h_{tt}}{2f_1}$,
\begin{align}
\begin{split}
\delta g_{tt} &= r^2 L^2 \delta h_{tt} + \mathcal{O}(r^0) \,, 
\\
\delta A_t &= r \frac{EL^2}{2}\frac{\delta h_{tt}}{f_1} + \mathcal{O}(r^0)\,.
\end{split}
\end{align}
Inserting this variation into~\eqref{E:deltaS1} and evaluating it on an AdS$_2$ solution~\eqref{E:electricAdS2} gives the dual stress tensor. This however identically vanishes,
\beq
\langle t^{tt}\rangle \equiv  \frac{2}{\sqrt{-h}}\frac{\delta S_1}{\delta h_{tt}} = 0\,,
\eeq
so the Virasoro symmetry at play has zero central charge.

However, this is not the end of the story. First, it is more natural to work in an alternative quantization, as one commonly does for Maxwell theory in AdS$_3$~\cite{Marolf:2006nd,Jensen:2010em}. The growing mode of $A_t$ ought to be fixed as a boundary condition, allowing the constant term to fluctuate. Second, the Cvetic and Papadimitriou~\cite{Cvetic:2016eiv} have argued that the renormalization scheme~\eqref{E:S1} is not quite correct. They advocate for the prescription
\begin{align}
\begin{split}
\label{E:SrenElectric}
S_{ren} &= \lim_{\Lambda\to\infty} \frac{1}{2\kappa^2}\left\{ \int d^2x \sqrt{-g} \Big( \varphi R + U - \frac{W}{4}F^2\Big) \right.
\\
& \qquad \quad+ \int dt \sqrt{-\gamma} \Big( 2\varphi K -L U \Big)
\\
& \qquad \quad - \left.\int dt\sqrt{-\gamma}\,W\Big( A_{\mu}F^{\mu\nu}n_{\nu}+ \frac{L}{4}F^2\Big)\right\}.
\end{split}
\end{align}
In alternate quantization we fix the canonical momentum conjugate to $A_t$, which in the bulk is
\beq
\pi^t \equiv \frac{\delta S_{bulk}}{\delta (\partial_r A_t)} = - \frac{WE}{2\kappa^2}\,.
\eeq
We identify this conjugate momentum as a fixed, external charge $\mathcal{Q}$ in the dual CQM.

This definition~\eqref{E:SrenElectric} has the virtue that it conserves the symplectic structure on the boundary.

Varying it with respect to the boundary metric gives
\beq
\langle t^{tt}\rangle = 0\,.
\eeq
The charge $\mathcal{Q}$ is conjugate to a $U(1)$ gauge field $\mathcal{A}$,
\beq
\langle \mathcal{A}\rangle =- \frac{\delta S_{ren}}{\delta \mathcal{Q}}\,.
\eeq
The charge $\mathcal{Q}$ is time-independent, and so $\langle \mathcal{A}\rangle$ is determined up to a total derivative, as is appropriate for a gauge field. That is, $\langle \mathcal{A}\rangle$ contains no local information. However, its holonomy around the Euclidean time circle is physical, encoding the chemical potential $\mu$.

To deduce the holographic formula for $\langle \mathcal{A}\rangle$ we need to deduce the variation of $S_{ren}$ with respect to $W[\varphi_0]E$. In order to remain on-shell, this variation induces a variation in the AdS radius and dilaton in accordance with~\eqref{E:familyOfAdS2}, with
\beq
\label{E:deltaQ}
\delta \mathcal{Q} = - \frac{1}{\kappa^2 E L^2}\delta \varphi\,.
\eeq 
The final result is that
\beq
\label{E:chemical}
\langle \mathcal{A}(t)\rangle = a(t)\,,
\eeq
for $a(t)$ the $\mathcal{O}(r^0)$ term in the Maxwell field~\eqref{E:electricAdS2}.

Now consider an electric AdS$_2$ black hole:
\begin{align}
\begin{split}
\varphi &= \varphi_0\,,
\\
A &= - EL^2 \left(r  - 2r_h + \frac{r_h^2}{r}\right)dt\,,
\\
g & =  L^2 \left( - r^2 \left( 1 - \frac{r_h^2}{r^2}\right)^2dt^2 + \frac{dr^2}{r^2}\right)\,.
\end{split}
\end{align}
We have chosen the $\mathcal{O}(1)$ term of the gauge field so that $A$ is regular at the Euclideanized horizon. The dual CQM is at a temperature $T = r_h/\pi$ and charge $Q = - WE/(2\kappa^2)$. The chemical potential is given by~\eqref{E:chemical} to be $\mu = 2\pi T E L^2$.

The Bekenstein-Hawking entropy is
\beq
S =\frac{2 \pi \varphi_0}{\kappa^2}\,.
\eeq
It can also be computed from the on-shell, Euclidean, holographically renormalized action, which is
\beq
\label{E:electricAdS2Sren}
S_{E} = \frac{2\pi \varphi_0}{\kappa^2} \,.
\eeq
The Euclidean action computes the canonical ensemble free energy $G(T,Q) = - T S_{E}$, whose variation gives the entropy and chemical potential as
\beq
\delta G = - S \delta T + \mu \delta Q\,.
\eeq
Using~\eqref{E:deltaQ} and taking thermodynamic derivatives, we indeed recover the Bekenstein-Hawking entropy and chemical potential $\mu = 2\pi T E L^2$.



\bibliography{refs}

\end{document}